\documentclass[12pt]{article}

\usepackage[dvips]{graphicx}

\begin{document}
\title{\textbf{Strong gravitational lensing of gravitational\\ waves in Einstein Telescope}}
\vspace{0.6 cm}
\author{Aleksandra Pi{\'o}rkowska$^1$, Marek Biesiada$^1$\footnote{Corresponding Author: marek.biesiada@us.edu.pl}, Zong-Hong Zhu$^2$}

\makeatletter

\def\@maketitle{%
  \newpage
  \null
  \vskip 2em%
  \begin{center}%
  \let \footnote \thanks
    {\LARGE \@title \par}%
    \vskip 1.5em%
    {\large
      \lineskip .5em%
      \begin{tabular}[t]{c}%
        \@author
      \end{tabular}\par}%
  \end{center}%
  \par
  \vskip 1.5em}

\makeatother

\maketitle

\begin{center}
$^1$ Department of Astrophysics and Cosmology, \\
Institute of Physics,University of Silesia, \\
Uniwersytecka 4, 40-007 Katowice, Poland\\
\vspace{0.2 cm}
$^2$ Department of Astronomy, Beijing Normal
University, \\Beijing 100875, China
\end{center}

\vspace{0.6 cm}

\begin{abstract}
\noindent

Gravitational wave experiments have entered a new stage which gets us closer to the opening a new observational window on the Universe. In particular, the Einstein Telescope (ET) is designed to have a fantastic sensitivity that will provide with tens or hundreds of thousand NS-NS inspiral events per year up to the redshift $z=2$. Some of such events should be gravitationally lensed by intervening galaxies.
\\
We explore the prospects of observing gravitationally lensed inspiral NS-NS events in the Einstein telescope. Being conservative we consider the lens population of elliptical galaxies. It turns out that depending on the local insipral rate ET should detect from one per decade detection in the pessimistic case to a tens of detections per year for the most optimistic case. The detection of gravitationally lensed source in gravitational wave detectors would be an invaluable source of information concerning cosmography, complementary to standard ones (like supernovae or BAO) independent of the local cosmic distance ladder calibrations.

\end{abstract}


\newpage

\section{Introduction} \label{sec:intro}

The last decade has brought a considerable development in design and operation of first laser interferometric
gravitational wave (GW) detectors (LIGO \cite{LIGO}, VIRGO \cite{VIRGO}, GEO-600 \cite{GEO} and TAMA-300 \cite{TAMA}). Despite the first scientific runs did not resulted in successful detections, the experience gained in these experiments (both in technology and data analysis) motivated people to upgrade the LIGO-VIRGO sensitivities to their advanced settings. Moreover a new generation detector called the Einstein Telescope (ET thereafter) is planned to be built and its design stage is currently being funded by the
European Framework Programme (FP7). Its sensitivity will be considerably improved over the existing detectors with expected detection rates from the inspiral NS-NS binaries as big as tens or hundreds of thousand events per year. For the details see the ET conceptual design study \cite{Abernathy03}.

In this paper, we calculate the predicted rate of strongly lensed gravitational wave sources to be seen by the Einstein Telescope. We restrict our predictions to double neutron star systems (NS-NS binaries)
because they comprise a known population of objects for which assessments of yearly detection rates are more reliable than for binary black hole systems (BH-BH) or mixed NS-BH binaries, even though these two last classes could have been seen from greater distances and in greater numbers. The study of BH binaries requires more thorough study of the formation history of BH-NS or BH-BH systems and will be a subject of a separate paper.

The paper is organized as follows. In Section~\ref{sec:rates}, we calculate the expected rate of NS-NS inspiralling systems for ET basing on the characteristics of the detector \cite{Abernathy03} in two settings: initial -- with noise curve approximated by polynomial expansion, and advanced -- the so called ``xylophone'' configuration. Then, in Section~\ref{sec:lensing} we discuss the strong lensing optical depth formalism for transient sources appropriate for NS-NS inspiralling systems to be observed in gravitational wave detectors. Section~\ref{sec:conclusions} contains the discussion of our results.

Throughout the paper we will assume flat FRW cosmological model as the one most supported by observations.
Recent results by PLANCK satellite created some tension (concerning value of the Hubble constant) with other studies ifinterpreted in the framework of $\Lambda$CDM model. However, we will rely on direct, locally well  calibrated studies \cite{Riess,Carnegie}. Consequently, the value of the Hubble constant will be assumed as: $H_0 = 74\; km/s\;Mpc^{-1}$ . If the unknown systematics in PLANCK experiment were ruled out, resolving the aforementioned tension would require a departure from the flat $\Lambda$CDM cosmology, in favour of a
non-trivial evolving dark energy equation of state. Therefore, we will phenomenologically describe the dark energy as a perfect fluid with barotropic equation of state $p = w(z) \rho$ with the $w$ coefficient given as $w = w_0 + w_a \frac{z}{1+z}$  -- the so called Chevalier-Polarski-Linder parametrization \cite{Chevalier01,Linder03}.

As it is well known, one can distinguish three types of distances in cosmology:

(i) proper distance:
\begin{equation} \label{proper}
r(z) = \frac{c}{H_0} \int _0^z \frac{dz'}{E(z')} =: \frac{c}{H_0} {\tilde r}(z)
\end{equation}

(ii) angular-diameter distance:
\begin{equation} \label{angular}
d_A(z) =  \frac{1}{1+z} \frac{c}{H_0} {\tilde r}(z)
\end{equation}

(iii) luminosity distance:
\begin{equation} \label{luminosity}
d_L(z) =  (1+z) \frac{c}{H_0} {\tilde r}(z)
\end{equation}

In the formulae above  we used a standard notation for an auxiliary quantity $E(z)$:
\begin{equation} \label{E}
E(z) = \sqrt {\Omega _m (1+z)^3 + (1 - \Omega _m) (1+z)^{3(1+w_0+w_a)} \exp \left(- \frac{3 w_a z} {1+z}\right) }
\end{equation}
Following the results of WMAP7 \cite{Komatsu2011} we assume the following values of parameters in the cosmic equation of state: $w_0 = -0.93$ and $w_a = -0.41$.

\section{Detection rate for unlensed events} \label{sec:rates}

Theory of gravitational wave detections has been reviewed many times in different papers (the original one being \cite{Finn93}). In particular we will refer to the one of most recent studies by \cite{TaylorGair}.

Matched filtering signal-to-noise ratio 
\begin{equation} \label{SNR}
\rho = 8 \Theta \frac{r_0}{d_L(z_s)} \left( \frac{{\cal M}_z}{1.2 M_{\odot}} \right)^{5/6}
\sqrt {\zeta(f_{max})}
\end{equation}
where: $d_L$ is the luminosity distance to the source, $\Theta$ is the orientation factor capturing part of sensitivity pattern due to (usually non-optimal) random relative orientation of NS-NS binary and the detector, $r_0$ is detector's characteristic distance parameter. It is estimated according to:
\begin{equation} \label{r0}
r_0^2 = \frac{5}{192 \pi} \left( \frac{3 G}{20} \right)^{5/3} x_{7/3} \frac{M_{\odot}^2}{c^3}
\nonumber
\end{equation}
where:
\begin{equation} \label{x7/3}
x_{7/3} = \int _0^{\infty} \frac{ df (\pi M_{\odot})^2}{(\pi f M_{\odot})^{7/3} S_n(f_{max})}
\nonumber
\end{equation}
Zeta parameter is defined as:
\begin{equation} \label{zeta}
\zeta(f_{max}) = \frac{1}{x_{7/3}} \int _0^{2f_{max}}\frac{ df (\pi M_{\odot})^2}{(\pi f M_{\odot})^{7/3} S_n(f_{max})}
\end{equation}

The distance parameter $r_0$ depends only on the detector's noise power spectrum. After \cite{TaylorGair} we consider two options: the polynomial approximation to the ET noise curve for its initial design, which gives $r_0 = 1527 \; Mpc$ and the advanced ``xylophone'' configuration, which gives $r_0 = 1918 \; Mpc$. For NS-NS inspiralling binaries, where the NS mass is close to the canonical value of $1.4\;M_{\odot}$ the maximal frequency (half of the orbital frequency at the end of inspiral) is:
\begin{equation}
f_{max} = \frac{785 Hz}{1+z_s} \left( \frac{2.8 M_{\odot}}{M} \right)
\end{equation}
$M$ here, is the total mass of the binary. It has been estimated (see e.g. \cite{TaylorGair}) that in this frequency range $\zeta(f_{max})$ factor is close to unity. Hence we assume from now on, that $\zeta(f_{max})=1$.

Let us denote  by ${\dot n}_0$ the local binary coalescing rate per unit comoving volume. We used several values of ${\dot n}_0$ motivated by population synthesis calculations of \cite{Abadie, Shaughnessy}. They are summarized in Table~\ref{n0}. Source evolution over sample is usually parametrized by
multiplying the coalescence rate by a factor $\eta(z)$, i.e. ${\dot n} = {\dot n}_0 \; \eta(z)$ where $\eta$ term can be approximated by a piecewise linear function \cite{Cutler, Schneider}:
\begin{eqnarray} \label{eta}
\eta(z) &=& 1 + 2\; z, \qquad {\rm if}\;\;\;
0 \leq z \leq 1 \\
\eta(z) &=& 0.75 \; ( 5 - z) \qquad {\rm if} \;\;\;
1 < z \leq 5  \\
\eta(z) &=& 0, \qquad {\rm otherwise} \nonumber
\end{eqnarray}
Such form is in agreement with simulations of merger rates based on population synthesis models made in \cite{TaylorGair} (their Fig.2).

\begin{table*}[ht]
\caption{Local NS-NS inspiral rate scenarios after \cite{Abadie}. Scenario called "Likely" taken after \cite{TaylorGair}.
Yearly detection rate calculated according to (\ref{Ndot}) in different scenarios is also given. } \label{n0}
\begin{center}  
\begin{tabular}{cccccc}

\hline
Inspiral rate & Max. & High & Reasonable & Likely & Low \\

\hline \\

${\dot n}_0$ [$Mpc^{-3}\;yr^{-1}$]  & $5.\;10^{-5}$ & $10^{-5}$ & $10^{-6}$ & $10^{-7}$ & $10^{-8}$ \\
\\
\hline

Yearly detection rate & Max. & High & Reasonable & Likely & Low \\

\hline \\

${\dot N}(>\rho_0)$ [$yr^{-1}$] &&&&& \\
initial design & $3.3\;10^6$ & $6.6\;10^5$ & $6.6\;10^4$ & $6.6\;10^3$ & $6.6\;10^2$ \\
``xylophone'' design & $6.\;10^6$ & $1.2\;10^6$ & $1.2\;10^5$ & $1.2\;10^4$ & $1.2\;10^3$ \\
\\
\hline

\end{tabular}\\
\end{center}
\end{table*}

The relative orientation of the binary with respect to the detector is described by the factor $\Theta$. This complex quantity cannot be measured nor assumed a priori. However, its probability density averaged over binaries and orientations has been calculated \cite{Finn93} and is given by a simple formula:
\begin{eqnarray} \label{P_theta}
P_{\Theta}(\Theta) &=& 5 \Theta (4 - \Theta)^3 /256, \qquad {\rm if}\;\;\;
0< \Theta < 4  \\
P_{\Theta}(\Theta) &=& 0, \qquad {\rm otherwise} \nonumber
\end{eqnarray}

\begin{figure}
\begin{center}
\includegraphics[angle=270,width=120mm]{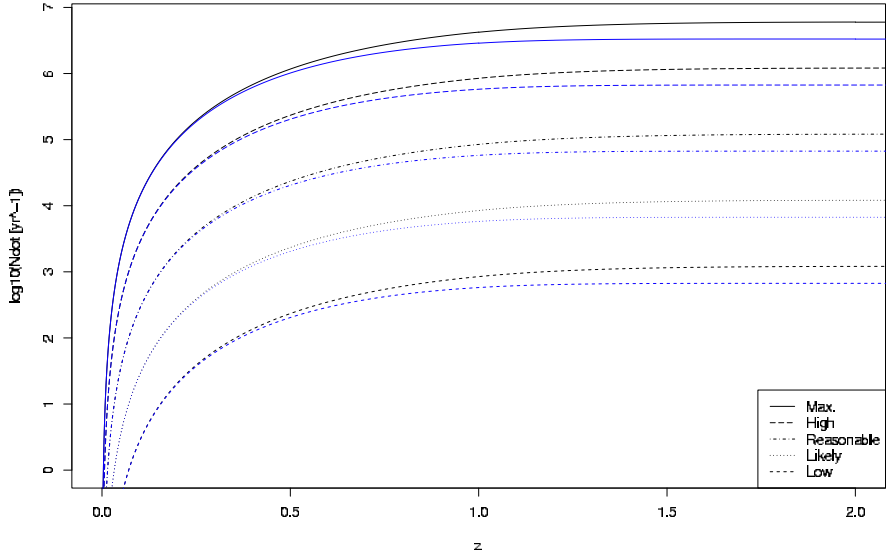}
\end{center}
\caption{\small{Yearly detection rate (as a function of source redshift) of NS-NS binaries by the Einstein Telescope for different local inspiral rate scenarios. Logarithmic scale (base 10) is adopted. Lower (blue) curves correspond to the initial ET sensitivity, upper ones to the advanced ``xylophone'' configuration.}
\label{Rate}}
\end{figure}

\begin{figure}
\begin{center}
\includegraphics[angle=270,width=120mm]{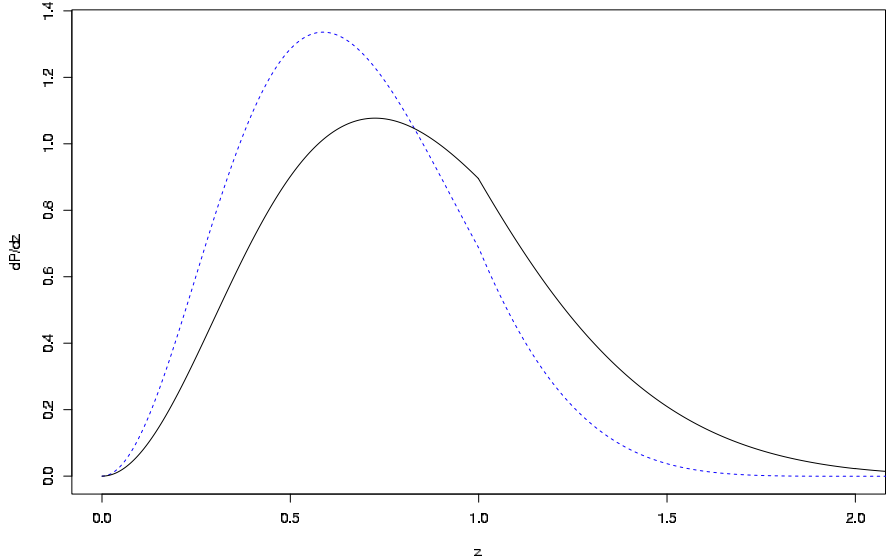}
\end{center}
\caption{\small{Probability density of NS-NS merging systems as a function of redshift. Dashed (blue) line corresponds to original ET design (polynomial approximation of the noise curve), solid (black) line corresponds to ``xylophone'' configuration.}
\label{Probability}}
\end{figure}

The rate $\displaystyle{\frac{d {\dot N}(>\rho_0)}{dz}}$ at which we observe the inspiral events (sources) that originate in the redshift interval $[z, \; z + dz]$ is given by \cite{Biesiada2001,Zhu2001}:
\begin{equation} \label{rate_nl}
\frac{d {\dot N}(>\rho_0)}{dz_s} = 4\pi \left( \frac{c}{H_0} \right)^3 \frac{\dot n_0}{1+z_s}\; \eta(z_s) \;  \frac{{\tilde r}^2(z_s)}{E(z_s)} \; C_{\Theta}(x(z_s))
\end{equation}
where $C_{\Theta}(x) = \int _x^{\infty} P_{\Theta}(\Theta) d\Theta$ denotes the probability that given detector registers inspiral event at redshift $z_s$ with $\rho > \rho_0.$  The quantity $C_{\Theta}(x)$ can be easily calculated as
\begin{eqnarray} \label{C_theta}
C_{\Theta} (x) &=& (1+x)(4-x)^4/256 \qquad {\rm for}\;\; 0\le x \le 4 \\
               &&   0        \qquad {\rm for}\;\;    x>4   \nonumber
\end{eqnarray}
where \cite{Finn93}:
\begin{equation} \label{x}
x(z) = \frac{\rho_0}{8.} (1+z)^{1/6} \frac{c}{H_0} \frac{{\tilde r}(z)}{r_0} \left( \frac{1.2\; M_{\odot}}{{\cal M}_0} \right)^{5/6}
\end{equation}
\noindent
From the equation (\ref{rate_nl}) one can calculate yearly detection rate of sources up to the redshift $z_s$:
\begin{equation} \label{Ndot}
{\dot N}(>\rho_0|z_s) = \int _0^{z_s} \frac{d {\dot N}(>\rho_0)}{dz} dz
\end{equation}
\noindent
In particular it is interesting to know the ${\dot N}(>\rho_0)$, i.e. the quantity (\ref{Ndot}) where $z_s$ is the limiting redshift corresponding to the detector's horizon. Figure~\ref{Rate} shows the expected detection rates of NS-NS inspiralling events to be seen by the Einstein Telescope for different local inspiral rate scenarios from Table~\ref{n0}. Probability density of inspiralling events as a function of source redshift is shown on Figure~\ref{Probability}. Let us notice that this probability density does not depend on ${\dot n}_0$ since it cancels in the normalizing factor.

\section{Lensing statistics} \label{sec:lensing}

The main focus of our paper is on lensing of distant GW sources therefore in this section we will present our approach which is similar to that of \cite{Sereno1,Sereno2} who considered predictions for LISA detector. We will make a conservative assumption that the population of lenses comprise only elliptic galaxies. Although there are known cases of spiral galaxies acting as lenses, the main factor influencing the lensing cross section is the mass. Consequently, the ellipticals created in mergers of low-mass spiral galaxies are more massive and indeed they dominate in all strong lensing surveys. We will model the lenses as singular isothermal spheres (SIS) which is surprisingly good approximation of early type galaxies acting as gravitational lenses \cite{Koopmans09}.

Characteristic angular scale of lensing phenomenon is set by the Einstein radius $\theta_E = 4 \pi \left( \frac{\sigma}{c} \right)^2 \frac{d_A(z_l,z_s)}{d_A(z_s)} $, where $\sigma$ is the velocity dispersion of stars in lensing galaxy, $d_A(z_l,z_s)$ and $d_A(z_s)$ are angular diameter distances between the lens and the source and to the source, respectively. If the source lies within the Einstein radius, i.e. if the angular position of the source with respect to the center of the lens is $\beta < \theta_E$, two images form on either side of the lens at (angular) locations $\theta_{\pm}= \theta_E \pm \beta$. It is convenient to use the Einstein radius as a unit: $x_{\pm} = \frac{\theta_{\pm}}{\theta_E}$, $ y = \frac{\beta}{\theta_E}$. Then the necessary condition for strong lensing (multiple images) is $y<1$, images form at $x_{\pm} = 1 \pm y$ with magnifications: $\mu_{\pm} = \frac{1}{y} \pm 1$.

Angular resolution of the Einstein Telescope (and any other gravitational wave detector) would be too poor to resolve lensed images (typical separations are of the order of $1''$). The only way to see gravitationally lensed GW signal is to register two time-delayed waveforms (coming form the same direction) having the same temporal structure (i.e. the frequency drift) but different amplitudes
\begin{equation} \label{waveform}
h_{\pm} = \sqrt {\mu_{\pm}} \; h(t) = \sqrt {\frac{1}{y} \pm 1}\; h(t)
\end{equation}
$h(t)$ here denotes the intrinsic amplitude (i.e. the one which would have been observed without lensing).
Time delay between images is equal to:
\begin{equation} \label{time delay}
\Delta t 
= \frac{32 \pi^2}{c} \left( \frac{\sigma}{c} \right)^4 \frac{d_A(z_l) d_A(z_l,z_s)}{d_A(z_s)} (1 + z_l) y =
\frac{32 \pi^2}{H_0} \left( \frac{\sigma}{c} \right)^4 \frac{{\tilde r}_l {\tilde r}_{ls}}{{\tilde r}_s} y = \Delta t_0 y
\end{equation}
It will be useful later on in the context of finite survey time. In order to be observed, the second image $x_{-}$ must be magnified above the threshold $\rho_0 = 8.$ In other words, its intrinsic
signal to noise ratio $\rho_{intr.}$ must be such that:
\begin{equation}
\sqrt {\frac{1}{y} - 1}\; \rho_{intr.} \ge 8.
\end{equation}
which translates into the demand that non-dimensional impact parameter $y$ should not exceed certain maximal value: $$ y \le y_{max} := \left[ 1+ \left( \frac{8.}{\rho_{intr.}} \right) ^2 \right] ^{-1} $$. In the rest of the paper we will limit ourselves to the sources with intrinsic signal-to-noise ratio equal to the threshold value of $\rho_{intr.} = \rho_0 = 8$, i.e. $y_{max} = 0.5$. This means that we neglect the magnification bias i.e. our estimates are even more conservative.

The elementary cross-section for lensing reads: $S_{cr} = \pi \theta_E^2 y_{max}^2$. Under the assumption of SIS lens we have:
\begin{equation}
S_{cr}   = 16 \pi^3 \left( \frac{\sigma}{c} \right)^4 \left( \frac{{\tilde r}_{ls}} {{\tilde r}_{s}} \right)^2 y_{max}^2
\end{equation}
Table~\ref{CrossSections} shows how much smaller are the lensing cross sections for sources intrinsically fainter than the threshold i.e. those neglected in further analysis.

\begin{table*}[ht]
\caption{\small{Ratio between elementary lensing cross sections for sources intrinsically fainter than the threshold and the threshold sources.}} \label{CrossSections}
\begin{center}  
\begin{tabular}{cc}

\hline
$\rho_{intr.}$ & $S_{cr}/S_{cr}(\rho_{intr.}=8)$  \\

\hline \\

3.  & 0.06  \\
4.  & 0.16  \\
5.  & 0.32  \\
6.  & 0.52  \\
7.  & 0.75  \\

\hline
\end{tabular}\\
\end{center}
\end{table*}

\begin{figure}
\begin{center}
\includegraphics[angle=270,width=120mm]{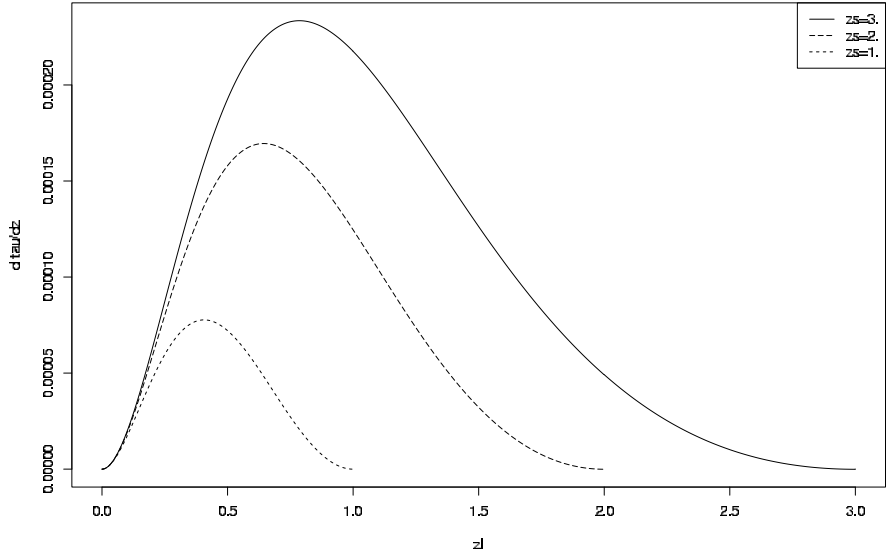}
\end{center}
\caption{\small{Differential optical depth for lensing as a function of lens redshift, shown for three sources of different redshifts.}
\label{Dtau}}
\end{figure}

Then it is straightforward to write a general expression for differential lensing probability by lenses at redshift in $[z_l, z_l + dz_l]$ with velocity dispersions in $[\sigma, \sigma + d \sigma]$:
\begin{equation} \label{diff tau}
 \frac{d^2 \tau}{d \sigma dz_l} = \frac{d n}{d \sigma} S_{cr}(\sigma, z_l, z_s) \frac{d V}{d z_l} = 4 \pi \left( \frac{c}{H_0} \right)^3 \frac{ {\tilde r}_l^2}{E(z_l)} S_{cr}(\sigma, z_l, z_s) \frac{d n}{d \sigma}
\end{equation}
where: $n(\sigma,z_l)$ denotes the comoving density of lenses and  $V(z_l)$ is the comoving volume. It is standard procedure to use the Schechter distribution function:
\begin{equation} \label{Schechter}
\frac{d n}{d \sigma} = n_{*} \left( \frac{\sigma}{\sigma_{*}} \right) ^{\alpha} \exp{ \left( - \left( \frac{\sigma}{\sigma_{*}} \right) ^{\beta} \right) }
\frac{\beta}{\Gamma (\frac{\alpha}{\beta}) } \frac{1}{\sigma}
\end{equation}
For the parameters entering the Schechter function we used the values obtained by \cite{Choi2007} from the SDDS DR5: $n_{*} = 8.\;10^{-3}\;(h/100)^3,$ $\sigma_{*} = 161 \pm 5\;\; km/s,$; $\alpha = 2.32 \pm 0.10$; $\beta = 2.67 \pm 0.07$. In principle one should have to account for the evolution of sources with redshift.
Despite there were some hints for the velocity dispersion evolution \cite{GavazziRuff}, we will assume no lens evolution. Figure~\ref{Dtau} displays differential optical depths for lensing evaluated according to (\ref{diff tau}) for sources lying at different redshifts.

Total optical depth for lensing up to the source redshift $z_s$ reads:
\begin{equation} \label{tau}
\tau = \frac{1}{4 \pi} \int _0^{z_s}\; dz_l \; \int ^{\infty}_0 \; d \sigma \; \frac{dn}{d \sigma} S_{cr}(\sigma, z_l, z_s) \frac{d V}{d z_l}
\end{equation}
In the case of non evolving lenses the integral over $\sigma$ can be done analytically leading to:
\begin{equation}
{\cal F}_4 = \int _0^{\infty} \left( \frac{\sigma}{c} \right) ^4 \frac{d n}{d \sigma} d \sigma = n_{*} \left( \frac{\sigma_{*}}{c} \right) ^4 \frac{\Gamma \left( \frac{4 + \alpha}{\beta} \right) }{\Gamma \left( \frac{\alpha}{\beta} \right)
}
\end{equation}
Then the differential cross section (w.r.t. the lens redshift $z_l$) reads:
\begin{equation} \label{dtaudz}
\frac{d \tau}{d z_l} = 16 \pi ^3 \left( \frac{c}{H_0} \right) ^3 \frac{{\tilde r}^2_{ls} {\tilde r}^2_l}{{\tilde r}^2_s E(z_l)} y_{max}^2 {\cal F}_4
\end{equation}
Now, bearing in mind that:
$$ \int_0^{z_s} \frac{{\tilde r}^2_{ls} {\tilde r}^2_l}{{\tilde r}^2_s E(z_l)} dz_l = \int_0^{{\tilde r}_s} \frac{{\tilde r}^2_{ls} {\tilde r}^2_l}{{\tilde r}^2_s } d {\tilde r}_l = \frac{1}{30} {\tilde r}_s^3 $$
one has:
\begin{equation}
\tau = \frac{16}{30} \pi ^3 \left( \frac{c}{H_0} \right) ^3 {\tilde r}_s^3 \left( \frac{\sigma_{*}}{c} \right)^4 n_{*} \frac{\Gamma \left( \frac{4 + \alpha}{\beta} \right) }{\Gamma \left( \frac{\alpha}{\beta} \right)} y_{max}^2
\end{equation}

Our considerations so far were strictly valid for infinitely long survey. If the survey lasts a finite time $T_{surv}$ some of the lensed transient signals (i.e. those occuring near the start or the end of the survey) would be missed because of the time delay (\ref{time delay}). Consequently the cross section for lensing should be modified appropriately (see e.g. \cite{Sereno2, Oguri}):
\begin{equation}
S_{cr} = 2 \pi \theta _E^2 \int _0^{y_{max}} f(\Delta t (y, z_l, z_s, \sigma))\; y\; dy
\end{equation}
Under assumption of uniform distribution of arrival times, one can write: $ f(\Delta t) = 1 - \frac{\Delta t}{T_{surv}} $, and:
\begin{equation} \nonumber
S_{cr} = \pi  \theta _E^2 \left( y_{max}^2 - \frac{2}{3} \frac{\Delta t_0}{T_{surv}} y_{max}^3 \right)
\end{equation}
Substituting this to (\ref{tau}), after calculating the integrals:
\begin{equation} \nonumber
{\cal F}_8 = \int _0^{\infty} \left( \frac{\sigma}{c} \right) ^8 \frac{d n}{d \sigma} d \sigma = n_{*} \left( \frac{\sigma_{*}}{c} \right)^8 \frac{\Gamma \left( \frac{8 + \alpha}{\beta} \right) }{\Gamma \left( \frac{\alpha}{\beta} \right)
}
\end{equation}

$$ \int_0^{z_s} \frac{{\tilde r}^3_{ls} {\tilde r}^3_l}{{\tilde r}^3_s E(z_l)} dz_l = \int_0^{{\tilde r}_s} \frac{{\tilde r}^3_{ls} {\tilde r}^3_l}{{\tilde r}^3_s } d {\tilde r}_l = \frac{1}{140} {\tilde r}_s^4 $$

and denoting: ${\cal F}_{*} = 16 \pi^3 {\cal F}_4$ and $\Delta t_{*} = \frac{32 \pi^2}{H_0} {\tilde r}_s \left( \frac{\sigma_{*}}{c} \right)^4 y_{max} $ one has:
\begin{equation} \label{tauDT}
\tau_{\Delta t} = \frac{{\cal F}_{*}}{30} \left( \frac{c}{H_0} \right) ^3 {\tilde r}_s^3 y_{max}^2 \left[ 1 - \frac{1}{7}
\frac{\Gamma \left( \frac{\alpha +8}{\beta} \right)}{\Gamma \left( \frac{\alpha +4}{\beta} \right)} \frac{{\Delta t}_{*}}{T_{surv}} \right] = \tau \left[ 1 -
\frac{1}{7}\frac{\Gamma \left( \frac{\alpha +8}{\beta} \right)}{\Gamma \left( \frac{\alpha +4}{\beta} \right)} \frac{{\Delta t}_{*}}{T_{surv}} \right]
\end{equation}

\section{Results and conclusions} \label{sec:conclusions}

\begin{figure}
\begin{center}
\includegraphics[angle=270,width=120mm]{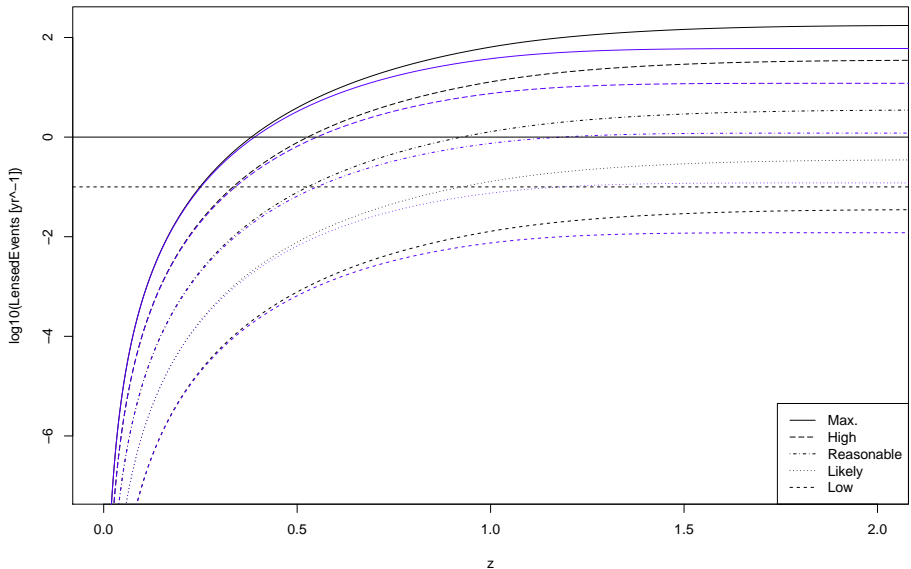}
\end{center}
\caption{\small{Expected strongly lensed events detection rate for different local inspiral rate scenarios. Logarithmic scale (base 10) is adopted. Lower (blue) curves correspond to the initial ET sensitivity, upper ones to the advanced ``xylophone'' configuration.}
\label{lensed}}
\end{figure}

Figure~\ref{lensed} shows the expected number of lensed inspiral GW events per year as a function of source redshift. As we have already discussed, in transient sources the optical depths for lensing depends on the survey time (see Eq.~\ref{tauDT}). Fig.~\ref{lensed} corresponds to the one-year scientific run of the ET. Horizontal solid line delimits ${\cal O}(1)$ detection rate, the dashed one corresponds to ${\cal O}(0.1)$ -- scenarios lying below the dashed curve have negligible chance of detecting strong lensing events.
One can see that in ``Max.'',``High'' and ``Reasonable'' scenarios of inspiral rates there are considerable chances that ET will register lensed GW inspiral events. Table~\ref{lensedTsurv} shows expected numbers of lensed NS-NS inspiral events for different survey lengths.

\begin{table*}[ht]
\caption{\small{Expected numbers of lensed GW events per year for different survey times.}} \label{lensedTsurv}
\begin{center}  
\begin{tabular}{ccccc}

\hline
ET configuration  & & & $T_{surv}$ &   \\
\hline
initial configuration & 1 year & 5 years & 10 years & continuous \\
\hline
Scenario & & & & \\
Max. & 60.1 & 64.0 & 64.5 & 65.0 \\
High & 12.0 & 12.8 & 12.9 & 13.0 \\
Reasonable & 1.2 & 1.3 & 1.3 & 1.3 \\
Likely & 0.06 & 0.1 & 0.1 & 0. 1 \\
Min & 0.006 & 0.01 & 0.01 & 0.01 \\
\hline
xylophone 	& 1 year & 5 years & 10 years & continuous \\
\hline
Scenario & & & & \\
Max. & 176.1 & 189.9 & 191.6	& 193.1 \\
High & 35.2 & 38.0 & 38.3 & 38.6 \\
Reasonable & 3.5 &  3.8 & 3.8 & 3.9 \\
Likely & 0.3 & 0.4 & 0.4 & 0.4 \\
Min	& 0.03 & 0.04 & 0.04 & 0.04 \\

\hline

\end{tabular}\\
\end{center}
\end{table*}

Our analysis was conservative as already stressed in respective places where conservative assumptions were made. However, the leading factor setting an order of magnitude to the strong lensing rate is the local inspiral rate of NS-NS systems. For optimistic scenarios, our analysis shows that there are good reasons to believe that the Einstein Telescope would register strongly lensed GW signals. If discovered, such systems would be of considerable importance. Time delays would have been measured with great accuracy. Identification of double image would be easy: they would appear as two time delayed signals having similar temporal structure (frequency drift) with different amplitudes. Amplitude ratio would allow to infer the source position $y$ \cite{Sereno2}, break the degeneracy and infer luminosity distance to the source $d_L(z_s)$. All this knowledge would be helpful in identification of lensing galaxy in the optical. Indeed, from potential candidates within the directional uncertainty of the ET one would have to search for massive galaxy at redshift range suggested by the distance ratio in time delay (\ref{time delay}) (assuming some standard value of the Hubble constant and typical velocity dispersion). Then if found, lensing galaxy could be a subject of more detailed spectroscopic studies leading in the end to precise measurement of the Hubble constant from time delay. Exploration of this possibility will be subject of a separate paper. In order to discuss it in a detailed manner one should properly account for external convergence (secondary lensing) \cite{Wambsganss}. Hopefully, there has recently been some progress in this field \cite{Amendola,Marra09,Marra10,Marra11}.

\section*{Acknowledgments}

The authors are grateful to the referee for his comments which allowed to improve the text substantially. 
A.P. and M.B. are partly supported by the Poland-China Scientific \& Technological Cooperation Committee Project No. 35-4. M.B. gratefully acknowledges hospitality in Beijing Normal University where this work was conceived. Z.-H.Z. is supported by the Ministry of Science and Technology National Basic Science Program (Project 973) under Grant No. 2012CB821804, NSFC Grant No. 11073005, the Fundamental Research Funds for the Central Universities, and the Scientific Research Foundation of Beijing Normal University.



\end{document}